\documentclass[a4paper]{mn2e}
\usepackage{epsf}

\title[Disc--black hole alignment]
{Aligning spinning black holes and accretion discs}

\author[A. R. King, S. H. Lubow, G. I. Ogilvie and J. E. Pringle]
{A. R. King$^1$, S. H. Lubow$^{2,3}$, G. I. Ogilvie$^2$ and
J. E. Pringle$^{1,2,3}$\\ $^1$Theoretical Astrophysics Group, University
of Leicester, Leicester LE1 7RH\\ $^2$Institute of Astronomy,
University of Cambridge, Madingley Road, Cambridge CB3 0HA\\ $^3$Space
Telescope Science Institute, 3700 San Martin Drive, Baltimore, MD
21218, USA}

\date{April 7, 2005}

\volume{000}

\pagerange{\pageref{firstpage}--\pageref{lastpage}} \pubyear{2005}

\begin{document}

\label{firstpage}

\maketitle

\begin{abstract}

We consider the alignment torque between a spinning black hole and an
accretion disc whose angular momenta are misaligned. 
%%%%%%%%%%%%%%%%%%
This situation must hold initially in almost all gas
accretion events on to supermassive black holes, and may occur in binaries
where the black hole receives a natal supernova kick.
%%%%%%%%%%%%%%%%%
We show that the torque always acts to
align the hole's spin with the total angular momentum without changing
its magnitude. The torque acts dissipatively on the disc, reducing its
angular momentum, and aligning it with the hole if and only if the
angle $\theta$ between the angular momenta ${\bf J}_d$ of the disc and
${\bf J}_h$ of the hole satisfies the inequality $\cos\theta >
-J_d/2J_h$. If this condition fails, which requires both $\theta >
\pi/2$ and $J_d < 2J_h$, the disc counteraligns.

\end{abstract}


\section{Introduction}

In a recent paper, Volonteri et al (2005; see also Madau 2004)
consider how the spins of supermassive black holes in galaxies change
as the holes grow through both mergers with other holes, and gas
accretion. Mergers occur at random angles, and when integrated over
the mass distribution expected in hierarchical models lead neither to
systematic spin--up nor spin--down. 
%%%%%%%%%%%%%%%%
Gas accretion, driven for example by minor mergers with satellite
galaxies, is likely also to occur at random angles, and thus be
initially retrograde with respect to the hole spin in half of
all cases.
%%%%%%%%%%%%%%%%
However, Volonteri et al. argue that gas
accretion nevertheless produces systematic spin--up, because a black
hole tends to align with the angular momentum of an outer accretion
disc on a timescale typically much shorter than the accretion
timescale for mass and angular momentum (Scheuer \& Feiler, 1996:
hereafter SF96; Natarajan \& Pringle, 1998).  Volonteri et al. (2005)
note that this conclusion holds only if most of the accretion takes
place through a thin accretion disc. If instead accretion is largely
via a geometrically thick disc (as happens if most mass accretes at
super-Eddington rates) alignment occurs only on the mass accretion
timescale. In this case there would be no net long--term spin--up,
assuming successive accretion events were randomly oriented.

Here we address the uncertainties in our current understanding of the
evolution of warped accretion discs, and the resulting uncertainties
in the alignment mechanisms and timescales for discs and black holes.
%%%%%%%%%%%%%%%%

We stress that
%%%%%%%%%%%%%%%
throughout the paper we neglect the change of the black hole spin as
it gains mass from the disc, i.e. we consider timescales shorter than
that for increasing the black hole mass significantly. Thus all the
torques we consider arise from the Lense--Thirring effect on a
misaligned disc. (These torques are dissipative, and can cause changes
in the local accretion rate in the disc. However these changes are at
most by factors $\sim 2$.)

We find that under some conditions counteralignment occurs, contrary
to what is usually thought. 
%%%%%%%%%%%%%%
As mentioned above, initial misalignment must characterise most
gas accretion events on to supermassive holes. Stellar mass black
holes accreting from a binary companion may also be misaligned in
cases where the hole received a supernova kick at formation.
%%%%%%%%%%%%%%
We begin by summarizing briefly what is currently known in various
cases.

\section{Dynamics of alignment}

\subsection{High viscosity discs}

The dynamics of the process of alignment of black hole and accretion
disc is not fully worked out. The best--understood case is when the
viscosity is sufficiently high and/or the disc sufficiently thin that
its tilt or warp diffuses through it in the way envisaged by Pringle
(1992). This is also the most likely case for black--hole discs in AGN
and binary systems (Wijers \& Pringle, 1999; Pringle, 1999). SF96
consider the case where a thin, high--viscosity disc is misaligned by
a small fixed angle $\theta$, and linearize in $\theta$. In this
approximation the diffusion equation governing the time evolution of
the disc surface density remains unchanged to first order, and the
analysis assumes that the disc has reached a steady state. In practice
this requires that we consider timescales longer than the inflow
timescale at (and within) some relevant radius $R_{\rm warp}$. SF96
(see also Rees, 1978) find that
\begin{equation}
R_{\rm warp} \sim \omega_p / \nu.
\label{rwarp}
\end{equation}
Here $\omega_p = 2GJ/c^2$, where $J = acM(GM/c^2)$ is the angular
momentum of the hole (with $-1 <a < 1$), and $\nu$ is the kinematic viscosity in the disc.
\footnote{SF96 assume that the effective viscosities in the disc,
$\nu_1$ and $\nu_2$ (Pringle 1992), are comparable.}

At radii $R \la R_{\rm warp}$ the spins of the disc elements are
aligned with the spin of the hole ${\bf J}_h$. At radii $R \ga R_{\rm
warp}$, the spins of the disc elements make an angle $\theta \ll 1$ to
${\bf J}_h$.  A spinning black hole induces Lense--Thirring precession
in the misaligned disc elements, and the precession rate falls off
rapidly with radius ($\propto R^{-3}$). Thus this induced precessional
torque acts mainly in the region around radius $R_{\rm warp}$. In
Cartesian coordinates let ${\bf J}_h$ define the $z$--axis and
consider an elemental annulus of the disc with spin ${\bf \Delta J}_d$
which is not parallel to ${\bf J}_h$.  Then each such disc annulus
feels a torque which tries to induce precession about the
$z$--axis. Integrating over all these torques to get the net torque on
the disc (and by Newton's third law the net torque on the hole) we
conclude that the net torque can only have components in the
$(x,y)$--plane. In the analysis of SF96 they assume that the disc
extends to infinite radius, and thus that the angular momentum of the
disc, ${\bf J}_d$, dominates that of the hole. If we assume that the
disc is tilted such that ${\bf J}_d$ lies in the $(x,z)$-plane, then
the conclusion of SF96 is that the $x$ and $y$ components of the
torque are equal in magnitude. The sign of the $y$ component, which
gives rise to precession, depends on the sign of the $z$ component of
${\bf J}_d$. The $x$ component affects the degree of misalignment of
the disc. It is negative both when the disc and hole are nearly
aligned (as above) and also when the disc and hole are nearly
counteraligned (i.e. when ${\bf J}_h \cdot\dot{\bf J}_d < 0$ ). This
implies that the net result is to try to align the spins of the disc
and the hole.  Thus the prediction of the SF96 analysis appears to be
that eventually ${\bf J}_h$ and ${\bf J}_d$ should end up parallel. We
see below that this conclusion does not hold in general.

\subsection{Low viscosity discs with small $a$ black holes}

If the disc is thick and/or its viscosity low, so that $\alpha < H/R$,
then its tilt or warp propagates as a wave rather than diffusing
(Papaloizou \& Lin 1995). If
such a disc is nearly aligned, relativistic precession effects do not
align its inner regions with the symmetry plane of the black hole, in
contrast to the viscous disc ($\alpha > H/R$). Instead the disc tilt
oscillates (Ivanov \& Illarionov, 1997; Lubow, Ogilvie \& Pringle,
2002) with an amplitude proportional to $R^{1/8}/(\Sigma H)^{1/2}$. In
the inner regions of black hole accretion discs this quantity
typically increases with decreasing radius (Shakura \& Sunyaev, 1973;
Collin--Souffrin \& Dumont, 1990). Thus even if the degree of
misalignment is small in the outer disc it can be large in the inner
disc, and the angular momentum vector of the matter actually accreted
at the horizon can make a large angle to the hole's spin. If the disc
and hole are close to being counteraligned, Lubow et al. (2002) show
that the inner disc and the hole align.

What interests us here are the disc torques on the hole. In Appendix
1 we summarise the particular analytic solution for the zero viscosity
case presented in Lubow et al (2002) and give explicit expressions for
the torques both when the disc and hole spins are nearly aligned, and
when they are nearly counteraligned. In the absence of viscosity the
torques result in mutual precesion of the disc and the hole. In
Appendix 2 we show how this analysis can be modified to take account
of a small viscosity, and we investigate the modified torques when
this has been introduced. In the nearly aligned case, we find that the
magnitudes of the torques are sensitive functions of the exact disc
parameters, because of the oscillatory nature of the disc tilt in the
inner regions. However, while the effect of this is to introduce great
uncertainty into the direction of the component of the torque which
gives rise to precession, we find that the sign of the component of
the torque which affects the angle between the spins of the hole and
the disc is exactly the same as in the viscous case analysis by
SF96.  We note that in practice this
small--viscosity case is most unlikely to hold for the alignment process
in most black--hole discs in AGN or X--ray binaries (Wijers \&
Pringle, 1999; Pringle, 1999) but that warp waves may be important in
the centres of these discs, where $H/R \sim 1$.

\section{Generalization}

Both sets of analysis reported above assumed that the degree of
misalignment was small (that is, the hole and disc were either nearly
aligned or nearly counteraligned) and that the disc tilt remained
fixed at large radius (that is, the angular momentum of the disc
dominates that of the hole).  We now consider the physics of the
general case where the angle of misalignment is not assumed to be
small, and the disc tilt is not assumed to be fixed. The angular
momentum of the hole ${\bf J}_h$ is well defined. We shall denote the
angular momentum of the disc as ${\bf J}_d$, but note that this is not
a well defined quantity. We discuss the exact meaning of ${\bf J}_d$
in this context below (Section 4.1). From these we construct a third
vector representing the total angular momentum ${\bf J}_t = {\bf J}_h
+ {\bf J}_d$, which is therefore a constant vector. The torques we are
interested in come about solely because there is a misalignment. We
now define the misalignment angle $\theta$ by
\begin{equation}
\cos \theta = \widehat{{\bf J}}_d . \widehat{{\bf J}}_h,
\end{equation}
where `hat' indicates a unit vector. We define $\theta$ so that $0 \le
\theta \le \pi$, with $\theta = 0 $ corresponding to full alignment
and $\theta = \pi$ corresponding to full counteralignment. The degree
of misalignment is measured by the vector ${\bf J}_d \wedge {\bf
J}_h$, and so any torques (which are vectors) must depend on this
quantity. Note that this vector is zero both for $\theta = 0$ and for
$\theta = \pi$. Then in the above discussion (Section 2.1), the $y$
axis is in the direction of ${\bf J}_h \wedge {\bf J}_d$, and the $x$
axis is in the direction ${\bf J}_h \wedge ({\bf J}_h \wedge {\bf
J}_d)$.

We have argued above, and indeed the analyses of both the high and low
viscosity cases confirm, that the torque on the hole cannot have a
component in the direction of ${\bf J}_h$. Thus the torque must have
the form
\begin{equation}
\frac{{\rm d} {\bf J}_h}{{\rm d}t} = - K_1 [ {\bf J}_h \wedge {\bf
J}_d ] - K_2 [ {\bf J}_h \wedge ({\bf J}_h \wedge {\bf J}_d) ].
\label{align}
\end{equation}
Here the first term through the quantity $K_1$ gives the magnitude and
sign of the torque which induces precession. It does not lead to a
change in $\theta$. The second term describes the torque which changes
the alignment angle $\theta$. Both sets of analysis in the high and
low viscosity cases show that $K_2$ is a positive quantity whose
magnitude is dependent on the properties of the disc and the hole.
%%%%%%%%%%%%%%%
Indeed we show below (\ref{dd}) quite generally that $K_2$ must be positive
in the presence of dissipation.
%%%%%%%
In
general of course $K_2$ is likely to be a function of $\theta$ as
well.

If we take the scalar product of this equation with ${\bf J_h}$ we see
that $dJ_h^2/dt = 0$, so that the magnitude of the spin of the hole
remains constant, i.e. $J_h$ = constant. Thus the tip of the ${\bf
J}_h$ vector moves on a sphere. The total angular momentum ${\bf J}_t
= {\bf J}_h + {\bf J}_d$ is of course a constant vector, representing
a fixed direction in space. Using this, and the fact that ${\bf J}_h
\cdot{\rm d}{\bf J}_h/{\rm d}t = 0$, we see that
\begin{equation}
{{\rm d}\over {\rm d}t}({\bf J}_h \cdot {\bf J}_t) = {\bf J}_t \cdot{{\rm
d}{\bf J}_h\over {\rm d}t} = {\bf J}_d \cdot {{\rm d}{\bf J}_h\over {\rm
d}t}.
\end{equation}
Using (\ref{align}) this leads to
\begin{equation}
{{\rm d}\over {\rm d}t}({\bf J}_h \cdot {\bf J}_t) = K_2[J_d^2J_h^2 -
({\bf J}_d\cdot {\bf J}_h)^2] \equiv A \ge 0.
\end{equation}
Now since both $J_h, J_t$ are constant, this means that 
\begin{equation}
{{\rm d}\over {\rm d}t}(\cos\theta_h) \ge 0 
\end{equation}
where $\theta_h$ is the angle between ${\bf J}_h$ and the fixed
direction ${\bf J}_t$. Thus $\cos \theta_h$ always increases, implying
that $\theta_h$ always decreases. This means that the angular momentum
vector of the hole
always aligns with the fixed direction corresponding to the total angular
momentum vector ${\bf J}_t$.

To see how ${\bf J}_d$ behaves during this process we consider the
quantity $A$ defined above (equation 5). We have, using the definition of ${\bf
J}_t$ and the fact that $J_h = $ constant,  that
\begin{equation}
A = {{\rm d}\over {\rm d}t}({\bf J}_h \cdot {\bf J}_t) = {{\rm
d}\over {\rm d}t}(J_h^2 + {\bf J}_h \cdot {\bf J}_d) = {{\rm d}\over {\rm
d}t}({\bf J}_h \cdot {\bf J}_d).
\end{equation}
Thus
\begin{equation}
{{\rm d}\over {\rm d}t}({\bf J}_h \cdot {\bf J}_d) = A \ge 0.
\label{hd}
\end{equation}
We have also that
\begin{equation}
0 = {{\rm d}\over {\rm d}t}J_t^2 = {{\rm d}\over {\rm d}t}(J_h^2 +
2{\bf J}_h \cdot {\bf J}_d + J_d^2),
\end{equation}
which implies that
\begin{equation}
{{\rm d}\over {\rm d}t}J_d^2 = -2 {\bf J}_d \cdot {{\rm d}{\bf J}_h\over {\rm
d}t} = -2A \le 0.
\label{dd}
\end{equation}

From this we conclude that the magnitude of the disc angular momentum
$J_d^2$ decreases as ${\bf J}_h$ aligns with ${\bf J}_t$. This is to
be expected since, although of course the total angular momentum of
the system (hole plus disc) is conserved, the alignment process requires
dissipation. Since the magnitude of the spin of the hole remains
unchanged, the dissipation must imply a reduction in (the magnitude
of) the disc angular momentum. 
%%%%%%%%%%%%%%%%%
This justifies our statement above that $K_2 > 0$, as a negative $K_2$
would require energy fed into the disc rotation.  Since we consider
timescales short compared with that for accretion this is not
possible.
%%%%%%%%%%%%%%%%

It is now straightforward to discover when the dissipative torque
leads to alignment or to counteralignment of ${\bf J}_h$ and ${\bf
J}_d$. Since the angle between ${\bf J}_h$ and ${\bf J}_d$ is
$\theta$, the cosine theorem gives
\begin{equation}
J_t^2 = J_h^2 + J_d^2 - 2J_hJ_d\cos(\pi - \theta)
\label{cos}
\end{equation}
Evidently counteralignment ($\theta \rightarrow \pi$) occurs if and
only if $J_h^2 > J_t^2$. This is equivalent to
\begin{equation}
\cos\theta < -{J_d\over 2J_h}.
\label{crit}
\end{equation}
Thus counteralignment is a possible outcome and requires
\begin{equation}
\theta > \pi/2,\ \  J_d < 2J_h. 
\label{cond}
\end{equation}

So why did the analysis of SF96, and also that given in the
Appendices, imply that the disc and hole always ended up aligned? The
reason is that both sets of calculations made the assumption that the
outer disc was fixed, that is that $J_d \gg J_h$. In this case we see
that counteralignment is forbidden, and alignment must result.

If instead $J_d < 2J_h$, then for ${\bf J}_h, {\bf J}_d$ in random
directions, counteralignment occurs in a fraction
\begin{equation}
f = {1\over 2}\biggl[1 - {J_d\over 2J_h}\biggr]
\end{equation}
of cases. The disc spin is given by the relevant root of (\ref{cos}),
i.e.
\begin{equation}
J_d = -J_h\cos\theta + (J_t^2 - J_h^2\sin^2\theta)^{1/2}
\label{da}
\end{equation}
for alignment, and
\begin{equation}
J_d = -J_h\cos\theta - (J_t^2 - J_h^2\sin^2\theta)^{1/2}
\label{dc}
\end{equation}
for counteralignment. In both cases $J_d$ decreases monotonically in
time, reaching the final values $J_t - J_h$ and $J_h - J_t$
respectively.

We can now derive the equation governing the change of $\theta$. From
(\ref{hd}) we have
\begin{equation}
A = J_h{{\rm d}\over {\rm d}t}(J_d\cos\theta) = J_hJ_d{{\rm d}\over {\rm
d}t}(\cos\theta) -A{J_h\over J_d}\cos\theta
\end{equation}
where we have used (\ref{dd}) to write ${\rm d}J_d/{\rm d}t =
-A/J_d$. Collecting terms and noting that $A =
K_2J_h^2J_d^2\sin^2\theta$ we have
\begin{equation}
{{\rm d}\over {\rm d}t}(\cos\theta) = K_2J_h\sin^2\theta(J_d +
J_h\cos\theta)
\label{rate}
\end{equation}
and from (\ref{da}, \ref{dc}) we get 
\begin{equation}
{{\rm d}\over {\rm d}t}(\cos\theta) = \pm K_2J_h\sin^2\theta(J_t^2 -
J_h^2\sin^2\theta)^{1/2}
\end{equation}
where $+/-$ corresponds to alignment/counteralignment
respectively. Expanding these two equations about $\theta = 0, \pi$
respectively shows that these equilibria are stable in the two
cases. Note that there is no contradiction between the global
alignment criterion (\ref{crit}) and the local equation (\ref{rate}):
$\theta$ does not always decrease monotonically for alignment or
increase monotonically for counteralignment.

\subsection{A geometrical picture}

Although in terms of the algebra given above the alignment process
looks somewhat complicated, in terms of the geometry of the situation
it is quite simple. 
\begin{figure}
  \centerline{\epsfxsize7cm \epsfbox{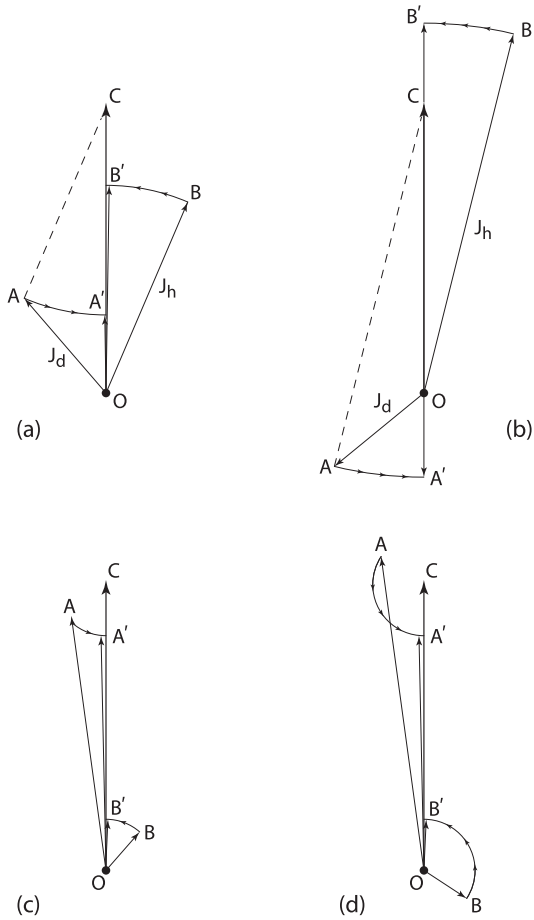}}
  \caption{The evolution of hole and disc angular momenta ${\bf J}_h
  $(OB),$ {\bf J}_d$ (OA) under the alignment torque, viewed in the
  plane they define. This plane precesses around the fixed total
  angular momentum vector ${\bf J}_t$ (OC). (a) A case where the
  initial angle $\theta$ between ${\bf J}_h, {\bf J}_d$ satisfies
  $\cos\theta > -J_d/2J_h$: the two angular momenta align. (b) A case
  where $\cos\theta < -J_d/2J_h$: the angular momenta
  counteralign. (c, d) Two cases where $J_d \gg J_h$ as considered by
  SF96 for which alignment always occurs.}
 \label{geom}
\end{figure}

In Figure~\ref{geom}(a) the initial vector ${\bf J}_d$ is represented
by the line OA, and the initial vector ${\bf J}_h$ by the line
OB. Then the total angular momentum ${\bf J}_t$ is represented by the
line OC, where OACB forms a parallelogram. Thus throughout the
subsequent evolution the line OC remains fixed. Since we are just
interested in the alignment process, rather than any precession around
${\bf J}_t$, we need only consider what happens in the plane defined
by OACB. In this plane we have seen that $J_h$ remains constant, and
that the effect of the evolution is to align ${\bf J}_h$ with ${\bf
J}_t$. Thus, as shown in the Figure, the tip B of the vector OB
describes the arc of a circle centred on O and ending up on
B$^\prime$. Once full alignment has occurred, the final vector ${\bf
J}_h$ lies along OB$^\prime$. Then in order that total angular
momentum be conserved, the tip A of the vector ${\bf J}_d$ must move
along a corresponding arc, centred on C, and ending at
A$^\prime$. Note that as this occurs, $J_d$ decreases
monotonically. We see that in this case (Figure~1(a)) the final vector
${\bf J}_d$ lies along OA$^\prime$, and the disc and hole end up
aligned.

In Figure~\ref{geom}(b) we show exactly the same procedure, but with different
initial values for ${\bf J}_d$ and ${\bf J}_h$.  As before the vector
${\bf J}_h$ moves from the initial position OB along an arc centred on
O to a final, aligned position, OB$^\prime$. The total angular
momentum ${\bf J}_t$, represented by OC, remains fixed. The disc
angular momentum vector OA moves along an arc, centred on C, to
its final position OA$^\prime$. But now, because of the initial values
of ${\bf J}_d$ and ${\bf J}_h$, the disc and the hole end up
counteraligned.

In Figures~\ref{geom}(c) and~\ref{geom}(d) we show the same evolution
but in the case $J_d \gg J_h$ considered in the analytic calculations
of SF96 and in the Appendix. We can see here that both disc and hole
always end up aligned, independent of the initial alignment of the
hole relative to the disc.

\subsection{Variation of $\theta$ with time}

\begin{figure}
  \centerline{ \epsfbox{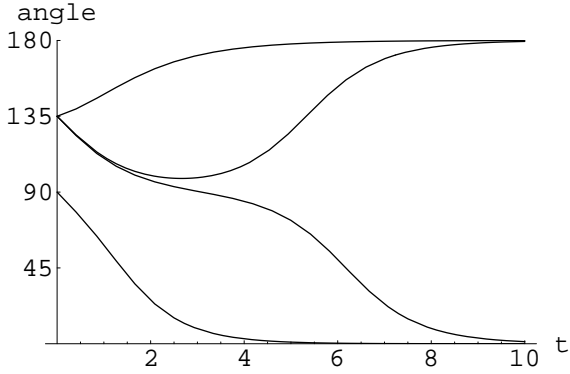}}
  \caption{The time evolution of the disc-black hole misalignment
 angle $\theta$ in degrees as a function of dimensionless time, which
 is normalized by $\tau$, the disc spin-down timescale for
 $\theta=90^o$. The evolution is determined by equation (\ref{rate})
 with an assumed constant value of $K_2 = 1/(\tau J_h^2)$.  Initial
 misalignment angles are $\theta = 135^o$ for the upper three curves
 and $90^o$ for the lowest curve.  The initial angular momentum ratios
 from the highest to lowest curves are $J_d/J_h = 0.5, 1.40, 1.42$ and
 0.5, respectively.  The curves, from the highest down, correspond to
 the cases a, b, d, and c, respectively in Fig~\ref{geom}.  For
 initial misalignment angle $135^o$, equation (\ref{crit}) predicts
 that the transition between long-term alignment and counteralignment
 occurs when initially $J_d/J_h = \sqrt{2} \approx 1.414$, as
 displayed in the middle two curves that are on opposite sides of the
 transition.  Notice that the second highest curve shows
 non-monotonic behaviour in time. }
 \label{theta-t}
 \end{figure}

Figure~\ref{theta-t} illustrates the evolution of misalignment angle
$\theta$ for a particular evolution rate,
namely for $K_2$ constant in equation (\ref{rate}).
The general behaviour of the solutions is independent
of the detailed form of $K_2$, provided that $K_2$ is
positive (i.e., $d J_d/dt <0$).

The highest and lowest curves in Figure~\ref{theta-t} show a monotonic
approach to misalignment and alignment, respectively.  The middle two
curves are close to the bifurcation between the two end states. From
equation (\ref{rate}), it follows that nonmonotonic behaviour in time
can occur for the case of a misaligned disc that begins an approach
toward alignment, but in the end becomes misaligned (as seen in the
second highest curve of Figure~\ref{theta-t}).  This situation is
realised for an initial state having $J_d < -2 J_h \cos \theta < 2
J_d$.  But in such cases, the disc never gets close to alignment; that
is, $\theta$ never drops below $\pi/2$.  A bifurcation between end
states of alignment/misalignment occurs for cases where initially $J_d
= -2 J_h \cos \theta$.  A misaligned disc that changes its sense of
evolution (i.e., the sign of $\dot{\theta}$) from heading toward
alignment to heading toward misalignment does so once $J_d$ has been
sufficiently reduced.  The extent of the required reduction is larger
for initial values of $J_d/J_h$ closer to the bifurcation value $-2
\cos{\theta}$, for which misalignment is achieved at a time when
$J_d(t)$ is zero.

\begin{figure}
  \centerline{\epsfxsize7cm \epsfbox{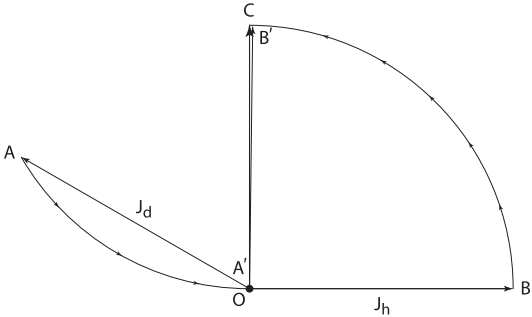}}
  \caption{ A geometrical picture of the critical case at the boundary
  between alignment and misalignment, following the notation of
  Figure~\ref{geom}.  In this case, we have $J_t = J_h$, with
  initially $J_d = -2 J_h \cos \theta$. The end state has ${\bf J}_h=
  {\bf J}_d$ and $J_d=0.$}
 \label{geom-crit}
 \end{figure}

Figure~\ref{geom-crit} illustrates the critical case of bifurcation,
according to the geometrical picture of the previous subsection.
Since $J_h = J_t$ in this case, we see that as ${\bf J}_h$ aligns
with ${\bf J}_t$, $J_d$ approaches zero.

Non-monotonic behaviour in the opposite sense from what is discussed
above, i.e., starting with an evolution toward alignment and ending
with misalignment, is only possible if $J_d$ increases in time. As
described in Section 4.2, this can happen because of the uncertainty
of what what is meant by $J_d$ in a real disc. Examples of such cases
are described in Section 4.2. These cases involve the complication
that $J_t$ also changes in time.

%%%%%%%%%%%%%%%%%%%  
\subsection{Comparison with numerical simulations}

To date there are two numerical simulations in the literature which
have $J_h > J_d/2$ and thus potentially allow the counteralignment we
predict. Nelson \& Papaloizou (2000) perform a quasi--Newtonian 3D SPH
simulation in which $J_h >> J_d$. They show that an initially prograde
disc does align with the hole, effectively also demonstrating again
that $K_2 > 0$ in the presence of dissipation. However they do not
consider any cases with an initially retrograde spin ($\theta >
\pi/2$). 

Fragile \& Anninos (2005) give fully relativistic 3D
grid--based simulations, again with $J_h >> J_d$. However there is no
explicit dissipation in their code ($K_2 \simeq 0$), so any
(counter)alignment can only occur on a long timescale associated with
numerical dissipation. There is indeed a hint of counteralignment in
the near--retrograde case they discuss.
%%%%%%%%%%%%%%%%%%

\section{Discussion}

We have considered the interaction between a misaligned accretion disc
and a rotating black hole. We have argued that the torque between them
must have the form (\ref{align}). The net result is twofold. First,
there is a component ($K_1$) which causes the disc and hole to precess
around the direction of the total angular momentum vector. The
direction and rate of precession can depend in a complicated way on
the properties of the disc. Second, there is a torque ($K_2$) which,
since $K_2 > 0$, acts to align the hole with the total angular
momentum without changing its spin rate.  This torque acts
dissipatively on the disc, and counteraligns or aligns it with the
hole according as the conditions (\ref{cond}) hold.

In the high viscosity case most relevant for black--hole discs the
steady disc shape is relatively simple (SF96). In both the co-- and
counter--rotating cases the disc is flat but inclined to the hole at
large radii, and flat but aligned with the hole at small radii. The
change between the two (the warp) occurs at a radius $R_{\rm warp}$
given by (\ref{rwarp}), where the rate at which the disc is twisted,
i.e. the Lense--Thirring precession rate $\omega_p /R^3$, is balanced
by the rate $\sim \nu/R^2$ at which viscous torques can propagate the
twist away. Here $\nu$ is the viscosity relevant to the process of
smoothing disc warp. It corresponds to the viscosity $\nu_2$,
introduced by Pringle (1992), which measures the viscosity
corresponding to the $(R, z)$--component of the stress tensor.

The actual dynamics of the various alignment processes are likely to
be complicated and need further investigation. As we have seen in
Section 3, the torque acts dissipatively on the disc, reducing its
angular momentum. If the total disc angular momentum is ${\bf J}_d$
then the disc eventually aligns with the hole if and only if the angle
$\theta$ between the angular momenta ${\bf J}_d$ of the disc and ${\bf
J}_h$ of the hole satisfies $\cos\theta > -J_d/2J_h$. If this
condition fails, which requires $\theta > \pi/2, J_d < 2J_h$, the disc
eventually counteraligns. This result just follows from the physical
nature of the torques, together with the fact that the process is
dissipative, so that $K_2 >0$. Based on these simple physical ideas we
were able to sketch the evolution of the two vectors ${\bf J}_h$ and
${\bf J}_d$. However, what we are not able to do, without further
consideration of the detailed properties of the disc in the form of
the coefficient $K_2$, is to predict the timescale on which this
happens.

\subsection{The meaning of ${\bf J}_d$}

So far we have been deliberately vague on the precise meaning of the
disc angular momentum ${\bf J}_d$. For an accretion disc we may define
the angular momentum vector ${\bf J}_d(R)$ of the material inside some
radius, $R$.  As an example we consider the disc model for AGN discs
given by Collin--Souffrin and Dumont (1990). For this disc model we are
interested in the innermost region (called Regime A, which corresponds
to region (b) in the disc models of Shakura and Sunyaev, 1973). In
this regime if we define the radius inside which the angular momentum
of the disc equals that of the hole as $R_J$, so that $J_d(R_J)
=  J_h$, then it is given in terms of the Schwarzschild radius of the
hole, $R_s$, as
\begin{eqnarray}
\label{RJ}
\lefteqn{\frac{R_J}{R_s} =
3.9 \times 10^3
\left(\frac{\epsilon}{0.1}\right)^{6/19}
\left(\frac{L}{0.1L_E}\right)^{-6/19}} \nonumber\\ 
\lefteqn{\ \ \ \ \ \ \times M_8^{-12/19}
\left(\frac{\alpha}{0.03}\right)^{8/19} a^{10/19}.}
\end{eqnarray}
Here $\epsilon$ is the efficiency of the accretion process (i.e. $L =
\epsilon \dot{M} c^2$), $L$ is the accretion luminosity, $L_E$ is the
Eddington limit, $M_8$ is the mass of the black hole in units of
$10^8$ M$_\odot$, $\alpha$ is the Shakura and Sunyaev (1973) viscosity
parameter, and $a$ is the (dimensionless) spin of the black hole.

The timescale on which this disc radius can communicate with the
central disc regions is the viscous timescale at this radius and is
given by
\begin{eqnarray}
\lefteqn{t_\nu(R_J) = 1.65 \times 10^8
\left(\frac{\epsilon}{0.1}\right)^{16/19}
\left(\frac{L}{0.1L_E}\right)^{-16/19}} \nonumber\\ 
\lefteqn{\ \ \ \ \ \ \ \ \times M_8^{6/19}
\left(\frac{\alpha}{0.03}\right)^{58/65} a^{14/19}\ {\rm yr}.}
\end{eqnarray}

Thus on timescales longer than this we expect the effective angular
momentum of the disc to dominate that of the hole and therefore that
on long timescales the spin of the hole ultimately aligns with that of
the disc as in Figures~\ref{geom}(c) and (d).

But on timescales less than this, the angular momentum of those parts
of the disc which are able to interact with the hole is much less that
that of the hole. On these shorter timescales we might expect the disc
evolution to resemble the evolution shown in Figures~\ref{geom}(a) and
(b). Thus we have the apparently contradictory possible scenario in
which on short timescales the disc tries to counteralign with the
hole, but on long timescales $t \gg t_\nu(R_J)$ it ends up co--aligning
with the hole. This means that the actual disc evolution depends
crucially on how the warp is propagated radially by the disc. In other
words we need to be able to predict the nature of the $(R,z)$--stress
denoted by the second viscosity $\nu_2$.

\subsection{Warp propagation}

If the degree of warping is very small compared to the disc thickness
$H/R$ then Papaloizou \& Pringle (1983) showed that, because of
resonant effects, the warp stress is much larger than the usual
azimuthal stress. If the warp stress is parametrised by $\alpha_2$ and
the usual viscosity by $\alpha_1$, then they found that $\alpha_2 =
1/(2 \alpha_1)$. In this case the warp radius can be quite small ---
$R_w/R_s \sim$ 10 -- 100 (Natarajan \& Pringle, 1998). However once
the warp becomes significant the approximations made in this analysis
break down. One possibility then is that the resonant flows become
unstable (Gammie, Goodman \& Ogilvie, 2000), the flow becomes
turbulent, and $\alpha_2$ is reduced significantly until perhaps
$\alpha_2 \sim \alpha_1$. If $\alpha_1 = \alpha_2$, which is the
assumption made by SF96, we find that
\begin{eqnarray}
\lefteqn{\frac{R_w}{R_s} = 990 \left(\frac{\epsilon}{0.1}\right)^{1/4}
\left(\frac{L}{0.1L_E}\right)^{-1/4} M_8^{1/8}} \nonumber \\
\lefteqn{\ \ \ \ \ \ \times \left(\frac{\alpha_1}{0.03}\right)^{1/8}
\left(\frac{\alpha_2}{0.03}\right)^{-5/8} a^{5/8}.}
\end{eqnarray}
In reality, it expected that disc viscosity is generated by
MHD turbulence, instigated by the magneto-rotational instability
(Balbus \& Hawley, 1991). How this turbulence interacts with a
rate of strain in the $(R,z)$--direction has yet to be fully determined
(see, for example, Torkelsson et al., 2000). It is evident that for
finite amplitude warps and misalignments, in order to estimate the
timescales and mechanisms for warp propagation it will be necessary to
undertake numerical simulations.

One of the first goals for such simulations will be to determine
whether the simple picture of two viscosities is adequate to a first
approximation (cf. Ogilvie, 2000) . Even in this picture it is clear
that a simple $\alpha$ prescription is inadequate. For example Larwood
et al. (1996) show that when a disc is subject to strong forced
precession (as is likely to occur in the inner regions of a tilted
disc around a Kerr black hole) the disc may break in the sense that
the disc tilt shows a sharp jump at some radius. This can only happen
in the diffusive picture if the diffusion coefficient ($\nu_2$) is a
function of the (gradient of the) disc tilt angle. But if this does
happen, it enhances the possibility, discussed above, of the inner
disc regions being able to counteralign on short timescales, before
eventually co-aligning at later times. If this happened, then
accretion onto the hole would act initially (on timescales $t \ll
t_\nu(R_J)$) to spin the hole down, in contradiction of the assumption
made by Volonteri et al. (2005).

\subsection{Black holes in X-ray binaries}

Maccarone (2002) reports that in at least two of the soft X--ray
transient (SXT) binaries (GRO J 1655-40 and SAX J 1819-2525) the
observed relativistic jets appear not to be perpendicular to the
orbital plane. If the jet directions are indicative of the direction
of the spin of the hole, then the most likely explanation is that the
misalignment occurred during the formation process of the black hole,
and that subsequent evolution has not had time to bring about
alignment. This interpretation is interesting in that it points to
black hole formation in a (presumably anisotropic) supernova
explosion.

From (\ref{RJ}) we see that for a $M \sim 10$ M$_\odot$ black
hole relevant for such binary systems the radius $R_J$ is typically
much larger than the binary separation. Thus in these systems $J_d \ll
J_h$. However, the angular momentum in the binary orbit is much larger
than that of the hole. Thus the crucial timescale in these systems is
the timescale on which tidal effects can transfer angular momentum
from the binary orbit to the disc.  On timescales shorter than this,
the evolution of the disc tilt follows that shown in Figures~\ref{geom}(a) and
\ref{geom}(b), with the possibility that the disc can counteralign with the
hole. Again, numerical simulations are required to provide estimates
of the tidal torques for strongly misaligned discs.

Estimates of timescales from stellar evolution theory can thus give
lower limits to the alignment timescales in SXT binaries. Maccarone
(2002) concludes that current theoretical estimates indicate that
alignment timescales are likely to be at least a substantial fraction
of the lifetimes of these systems. In any case the long quiescent
intervals (10 -- 50~yr or more) in SXT binaries strongly suggest that
the inner regions of the disc are either absent or very tenuous. This
means that virtually all of the disc mass is far outside the warp
radius ($\sim$ a few Schwarzschild radii) and so the alignment torque
must be very weak.

\section{Acknowledgments} 

ARK acknowledges a Royal Society--Wolfson Research Merit Award. JEP
thanks STScI for continued support under their Visitor Program.
%%%%%%%%%%%%%%%%%%
We thank the referee for helpful remarks concerning numerical
simulations.
%%%%%%%%%%%%%%%%%%

\bigskip
\noindent
{\bf Appendix 1: zero viscosity discs with small $a$ black holes}

Lubow et al. (2002) consider the torque exerted between the central
black hole and a disc for the case of zero viscosity. They use the
same linearised approximation as SF96, with the disc again assumed
steady. At each radius the disc angular momentum is in the direction
of the unit vector ${\bf l}(R) \approx (l_x, l_y, 1)$, where $l_x, l_y
\ll 1$. They then use the complex quantity $W(R,t) = l_x + i l_y$ to
describe the shape of the disc. They consider explicitly a simple
example in which the amplitude of oscillations is independent of
radius, and for which there is a simple analytic solution for $a^2 \ll
r = R/(GM/c^2)$. They take the disc thickness to vary as
\begin{equation}
\frac{H}{R} = \epsilon r^{h-1},
\end{equation}
and the disc surface
density to vary as
\begin{equation}
\Sigma = \Sigma_o r^{\frac{1}{4} - h},
\end{equation}
where we require $h > -\frac{1}{8}$.

In this case, if we set
\begin{equation}
x = \left( \frac{24 \mid a \mid }{\epsilon^2} \right)^{1/2}
\frac{r^{-(h+1/4)}}{h+ \frac{1}{4}},
\end{equation}
where $a$ is the black hole spin parameter with positive (negative)
$a$ corresponding to alignment (counteralignment).

If $a > 0$, Lubow et al. (2002) find that
\begin{equation}
W = W_\infty \frac{\cos(x_{\rm in} - x)}{\cos x_{\rm in}},
\end{equation}
where $W_\infty$ gives the tilt at large radius, $x_{\rm in}$
corresponds to the inner boundary where the torque vanishes, i.e.
$dW/dr =0$, and we need the {\it proviso} that $\cos x_{\rm in} \neq
0$.

If $a < 0$, it is simple to show that the corresponding expression is
\begin{equation}
W = W_\infty \frac{\cosh(x_{\rm in} - x)}{\cosh x_{\rm in}}.
\end{equation}

If $T_x$ and $T_y$ are the components of the torque on the disc,
assuming as above that the hole spin is aligned along the $z$-axis,
and writing $T = T_x + i T_y$, we find that

\begin{equation}
T = \frac{\sqrt{6} \, i \, \pi}{3} |a|^{1/2} \epsilon \frac{G^2 M^2
 \Sigma_o}{c^2} \int_0^{x_{\rm in}} W \, dx.
\end{equation}

Then in the case we are considering for $a > 0$
\begin{equation}
T \propto i \mid a \mid^{1/2} W_{\infty} \tan x_{\rm in},
\end{equation}
and for $a < 0$
\begin{equation}
T \propto - i \mid a \mid^{1/2} W_{\infty} \tanh x_{\rm in}.
\end{equation}
In both cases the coefficient of proportionality is real and positive.
Thus, in this case with zero disc viscosity the torque causes a mutual
precession on the hole and the disc. But unlike in the high viscosity
case, if $a > 0$ the sign of the precession is determined by the
details of the disc properties.

\bigskip
\noindent
{\bf Appendix 2: small viscosity discs with low $a$ black holes}

It is not possible to introduce a small constant $\alpha$ viscosity as
a small perturbation of the Lubow et al. analysis. This is because at
large radii, the Lense-Thirring effect goes to zero, and so the small
viscous perturbation does not remain small. But, one can apply
perturbation theory for a spatially varying $\alpha$ that takes the
form
\begin{equation}
\alpha = \alpha_o r^{-1}
\end{equation}
for the case of small viscosity.

If we write
\begin{equation}
\zeta = \surd ( 1 + \frac{i}{3} \alpha_o ),
\end{equation}
where it is understood that we take the root with positive
real part, then in the
particular case considered above the solutions become for $a > 0$
\begin{equation}
W = W_\infty \frac{ \cos [ \zeta (x_{\rm in} -x)]}{\cos (\zeta x_{\rm
in} ) },
\end{equation}
and for $a < 0$,

\begin{equation}
W = W_\infty \frac{ \cosh [ \zeta (x_{\rm in} -x)]}{\cosh (\zeta x_{\rm
in} ) }.
\end{equation}

\begin{figure}
  \centerline{\epsfbox{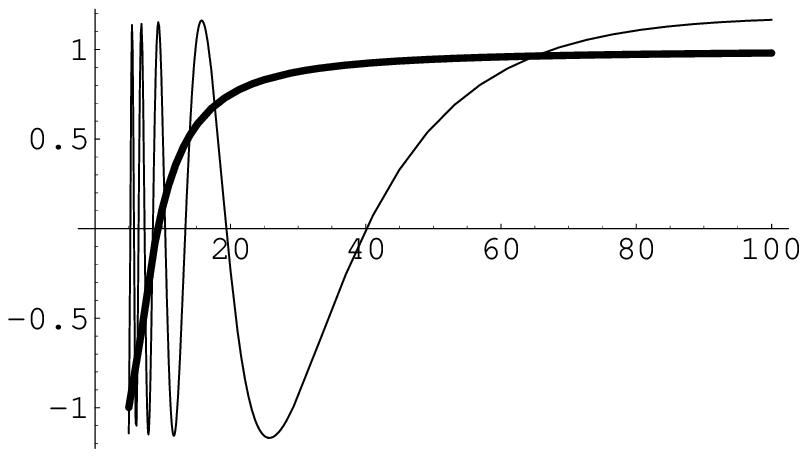}}
  \centerline{\epsfbox{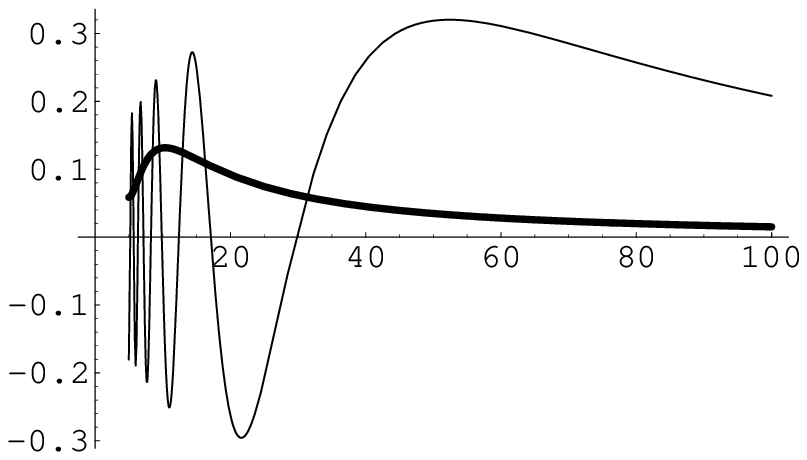}}
  \caption{The real (top) and imaginary (bottom) parts
    of $W(r)/W_\infty$ as a function of radius $r$ (in units of $G
    M/c^2$) for two disc-black hole cases. In both cases, we adopt
    $a=0.3$, and $h=1$. The inner radius $ r_{\rm in}$ occurs at the
    marginally stable orbit.  The oscillatory (wave-like) solutions
    are for $\epsilon = 0.01$, while the non-oscillatory solutions
    (plotted with the heavier lines) occur for $\epsilon=0.1$.  The
    viscosity parameter $\alpha= \epsilon r_{\rm in}/r$.}
    \label{W-r}
\end{figure}

The warps acquire a $y-$component that is out of phase with respect to
$W_\infty$ as a consequence of the viscosity. This phase shift leads
to a net alignment torque (nonzero $x$-component of torque) on the
disc.  In the $a >0$ case, radially oscillatory warped waves can occur
provided that $x_{\rm in} > 2 \pi$. Such waves are possible for discs
that are sufficiently thin ($\sqrt{a}/\epsilon$ sufficiently large).
In Figure~\ref{W-r}, we plot  $W$ as a function of $r$ for two sets of disc
parameters that differ in the value of $\epsilon$. Notice that
oscillatory behaviour occurs for the thinner disc.

The torque on the disc is given by
\begin{equation}
\label{T}
T = \frac{\sqrt{6}\, i \, \pi}{3} a^{1/2} \epsilon \, W_{\infty}
  \frac{G^2 M^2 \Sigma_o}{c^2} \frac{\tan [ \zeta x_{\rm in}]}{ \zeta}
  .
\end{equation}
This torque applies to both positive and negative values of $a$
through analytic continuation.  It is straightforward to show that the
$x$-component (real part) of the torque is negative, independent of
the sign of $a$.  This result implies that the degree of misalignment
decreases in time, as will be discussed more fully in the next
subsection.

We consider discs with inner truncation at the radii of the marginally
stable orbits.  In the prograde spin case ($a>0$), the alignment
torque $Re(T)$ can undergo large variations as a function of
parameters $a$ and $\epsilon$. But, there is a well defined average
value.  

We determine the $a$-averaged value of the alignment torque in the
case of $a>0$ with a fixed value of $\epsilon$.  For $a \gg
\epsilon^2$ (i.e., $x_{\rm in} \gg 1 $), the torque undergoes multiple
local peaks in value where $x_{\rm in}(a) = n \pi/2$ for positive
integer $n$. Near such points,
\begin{equation}
\label{tanap}
\tan [ \zeta x_{\rm in}] \simeq \left(c_1 \frac{a^{1/2}}{\epsilon}
-\frac{n \pi}{2} + i c_2 n \alpha_o \right)^{-1},
\end{equation}
where $c_i$ are real constants of order unity and we have ignored
variations in the inner disc radius (marginally stable orbit) as a
function of $a$.  Since these peaks are spaced in $a$ by an amount
$c_3 n \epsilon^2$, the $a$-averaged value of $q(a)=Im(\tan [ \zeta
x_{\rm in}(a)])$ can then be expressed as
\begin{equation}
<q(a)> = -\int_{-w}^w \frac{dz}{z^2+1}
\end{equation}
where $z=(c_1 a^{1/2}/\epsilon -n \pi/2)/(n \alpha_o c_2)$.  $w$ is
the peak width expressed in terms of $z$ which is inversely
proportional to $\alpha_o$.  For small $\alpha_o$, we can take the
integral limits to infinity and we find that $q$ is independent of
$\epsilon$ and $\alpha_o$.  Consequently, we can approximate the
average alignment torque for $a>0$ by taking the average value of
$\tan [ \zeta x_{\rm in}]/ \zeta$ in equation (\ref{T}) to be a
constant of order unity that is independent of $\epsilon$ and
$\alpha_o$.

For a fixed value of $\epsilon$ such that $x_{\rm in} \gg1$
(so that wave-like behaviour occurs), the $a$-averaged value of the
torque from 0 to $a$ is approximately
\begin{equation}
\label{Tp}
<Re(T)> \approx - a^{1/2} \epsilon \, W_{\infty} \frac{G^2 M^2 \Sigma_o}{c^2}.
\end{equation}
This average torque is then independent of viscosity parameter $\alpha_o$.
The precessional torque is of similar order.

\begin{figure}
  \centerline{\epsfbox{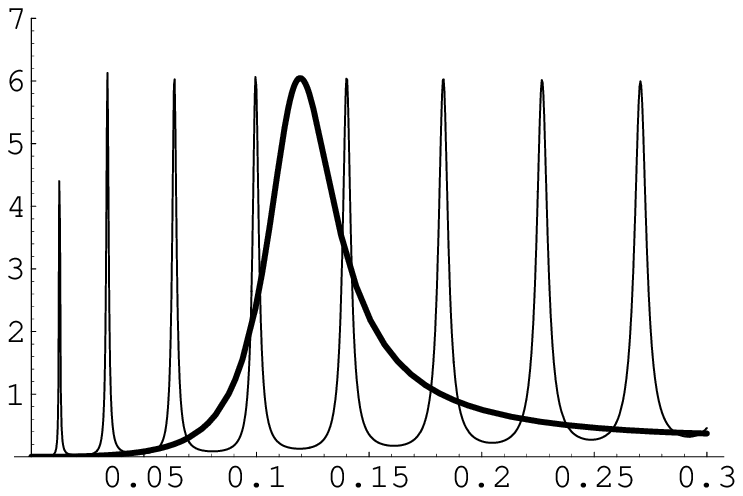}}
  \centerline{\epsfbox{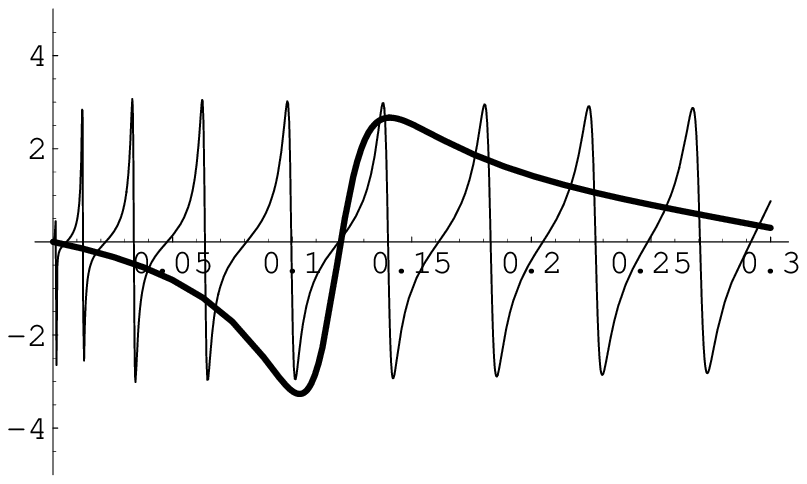}}
  \caption{The  negative dimensionless alignment
    torque (top) and the dimensionless precession torque (bottom) on
    the disc as a function of $a>0$. The dimensional torques are
    recovered by multiplying by $\epsilon \, W_{\infty} \, G^2 M^2
    \Sigma_o/c^2$.  The disc inner radius $ r_{\rm in}$ occurs at the
    marginally stable orbit for each $a$. Parameter $h$ is unity.  Two
    cases of disc-black hole systems are plotted, corresponding to the
    two cases in Figure 3 (but with varying $a$).  The more strongly
    fluctuating torques are for $\epsilon = 0.01$, while the smoother
    torques (plotted with the heavier lines) occur for $\epsilon=0.1$.
    The viscosity parameter $\alpha= \epsilon r_{\rm in}/r$.  }
    \label{ta}
\end{figure}

In Figure \ref{ta}, we plot the dimensionless torques on the black hole as a
function of $a>0$.  Notice that the disc alignment torque is negative,
indicating that alignment occurs. The precession rates undergo changes
in sign as the spin rate $a$ changes.
 
For the retrograde spin case ($a < 0$) with $x_{\rm in} \gg 1$, the
torque follows from equation (\ref{T}),
\begin{equation}
\label{T-ret}
 T = -\frac{\sqrt{6 }\, \pi}{3} \left( i + \frac{\alpha_o}{6} \right)
\epsilon |a|^{1/2} W_{\infty} \frac{G^2 M^2 \Sigma_o}{c^2}.
\end{equation}
In this case, we see that the ratio of the alignment torque to the
precessional torque is $\alpha_o/6$. Furthermore, the precession does not
change direction as a function of $a$, as was found in the case for
$a>0$ (see Fig \ref{ta}). Consequently, for low values of the turbulent
viscosity parameter, the alignment time scale can be much longer
than the precession time scale. This situation is unlike the case for
$a>0$, where both time scales are comparable.

For both the prograde or retrograde cases, the torque on the disc is
exerted where $x \simeq1$. The torque radius in units of $G M/c^2$ is
then given by
\begin{equation}
 \label{rT}
 r_{\rm T} = \left(
 \frac{
 \sqrt{24 |a|}
 }
 {\epsilon}
 \right)^{\frac{1}{h+\frac{1}{4}}}.
\end{equation}
Consequently, for the purposes of computing torques, such as in
equation (\ref{T-ret}), the viscosity parameter $\alpha_o$ is related to
$\alpha = \alpha_o/r$ by
\begin{equation}
\alpha_o \approx  \frac{\alpha \sqrt{|a|}}{\epsilon},
\end{equation}
for $h \approx 1$.

\label{lastpage}

\end{document}